\documentclass[aps,pra,twocolumn,showpacs,superscriptaddress]{revtex4}
\usepackage{graphicx}
\usepackage{amsmath}
\usepackage{amssymb}

\input{epsf}
\begin{document} 

\title{
Modification of roton instability due to the presence of a second dipolar Bose-Einstein condensate
}

\author{M. Asad-uz-Zaman}
\affiliation{Department of Physics and Astronomy,
Washington State University,
Pullman, Washington 99164-2814, USA}
\author{D. Blume}
\affiliation{Department of Physics and Astronomy,
Washington State University,
  Pullman, Washington 99164-2814, USA}

\date{\today}

\begin{abstract}
We study the behavior of two coupled purely dipolar Bose-Einstein condensates, each located in a cylindrically symmetric pancake-shaped external confining potential, as the separation $b$ between the traps along the tight confining direction is varied. The solutions of the coupled Gross-Pitaevskii and Bogoliubov-de Gennes equations, which account for the full dynamics, show that the system behavior is modified by the presence of the second dipolar BEC. For sufficiently small $b$, the presence of the second dipolar BEC destabilizes the system dramatically. In this regime, the coupled system collapses through a mode that is notably different from the radial roton mode that induces the collapse of the uncoupled system. Finally, we comment on the shortcomings of an approach that neglects the dynamics in the $z$-direction, which is assumed to be a good approximation for highly pancake-shaped dipolar BECs in the literature.
\end{abstract}
  

\maketitle

\section{Introduction}
\label{introduction}

Dipole-dipole interactions are long-range and anisotropic and dominate the behavior of many liquids and solids such as ferrofluids and superfluid $^3$He~\cite{laha07,bara08,laha09,carr09,ghoh97}. The condensation of $^{52}$Cr atoms~\cite{grie05}, which have a large magnetic dipole moment compared to alkali atoms, paved the way for studying the physics of long-range interactions in a clean model system~\cite{laha07,grie05,giov06,bism10}. While the dynamics of dipolar Bose-Einstein condensates (BECs) is, in general, governed by an interplay between the short-range $s$-wave interactions and the long-range dipole-dipole interactions, the $s$-wave scattering length can be tuned to vanish through the application of an external magnetic field in the vicinity of a Fano-Feshbach resonance~\cite{wern05}. This feature allows for the experimental realization of purely dipolar BECs. The present work investigates the behavior of two coupled dipolar BECs in a double-well type set up within the mean-field framework, which is expected to describe the key features of dipolar gases such as Cr BECs properly but not necessarily those of molecular samples such as RbK~\cite{ni08,bara08,laha09,carr09}. Previous mean-field studies of single dipolar BECs in a pancake-shaped external trap predicted interesting features such as a red blood cell type shaped ground state density as well as collapse induced by radial and angular roton modes~\cite{dutt07,rone07,wils09}.

The behavior of dipolar BECs is even richer when a double well geometry is considered~\cite{xion09,asad09,abad10}. Describing the condensate by a single mean-field wave function, the existence of an instability island immersed in an otherwise stable region has been predicted to exist for certain parameter combinations~\cite{asad10}. Furthermore, macroscopic quantum self trapping, a phenomenon intensely studied for $s$-wave interacting BECs~\cite{smer97,ragh99,gati07}, has been predicted to occur for a dipolar BEC in a cigar shaped double-well potential~\cite{asad09,xion09}. The transition from the macroscopic quantum self-trapping to the Josephson oscillation regime has been interpreted using a single two-mode model that treats the left well and the right well as being occupied by macroscopic wave functions $\Psi_1$ and $\Psi_2$, respectively. Extensions to triple-well potentials, which provide a simplifying model of an optical lattice system, have also been considered~\cite{laha10}. Here, we model a two-well dipolar system, for which tunneling is assumed to be negligible, and solve a set of two coupled Gross-Pitaevskii (GP) and Bogoliubov-de Gennes (BdG) equations. Unlike the two-mode model eluded to above and unlike related earlier studies~\cite{kobe09,wang08,klaw09}, our approach accounts for the full system dynamics within the mean-field framework. The dipoles in the two traps are assumed to be aligned along the tight confining direction and the system behavior is investigated as a function of the separation between the two clouds.  

The remainder of the paper is organized as follows. Section~\ref{sec_theory} introduces the stationary and dynamical mean-field description of two coupled dipolar BECs. Section~\ref{sec_results} presents and interprets our numerical results of the stability of the system as functions of the dipole strength, the aspect ratio and the separation between the two clouds. Lastly, Sec.~\ref{sec_conclusion} concludes.

\section{Mean-field description}
\label{sec_theory}
\subsection{Coupled Gross-Pitaevskii equations}
\label{sec_gpe}

 We consider two dipolar systems, each confined by a cylindrically symmetric external trap $V_{{tj}}(\rho, z)$, $j = 1$ and $2$,
\begin{equation}
\label{eq_trap}
V_{tj}(\rho,z) = \frac{1}{2} m \omega_{\rho}^2 \big(\rho^2 + \lambda^2 \left(z-z_{j0}\right)^2\big)
\end{equation}
with aspect ratio $\lambda$, where $\lambda = \omega_z / \omega_{\rho}$, and $\omega_{\rho}$ and $\omega_{z}$ denote the angular trapping frequencies along the $\rho$ and $z$ directions, respectively. Here, we use cylindrical coordinates [$\vec{r}$ =($\rho$,$\phi$,$z$)]. In Eq.~(\ref{eq_trap}), $m$ denotes the mass of the dipoles and the $z_{j0}$ denote the trap centers along the $z$-direction. We consider the same type of atomic species in both traps, e.g., $^{52}$Cr in the same internal state. Throughout, we assume that the number $N_j$ of dipoles in the $j^{th}$ trap ($j = 1$ or $2$) is fixed, i.e., we assume that tunneling between the traps is absent. Our main interest is in determining the system behavior for aspect ratios of the order of $10$ as the distance $b$, $b = |z_{10} - z_{20}|$, is varied. The parameter $b$ determines the effective coupling between the two dipolar BECs. When $b$ is infinitely large, the effective coupling vanishes and the system behaves like two independent dipolar BECs. When $b$ is small, the effective coupling is strong and the system behavior is changed due to the long-range and anisotropic dipole-dipole interaction between the dipoles located in the  two traps. Throughout, we assume that the dipoles are aligned along the $z$-direction, so that the dipole-dipole interaction potential is given by $V_{\mathrm{dd}}(\vec{r})=d^2 \frac{1 - 3 \cos^2 \vartheta}{r^3}$, and  $s$-wave interactions are neglected. Here, $d$ is the dipole strength, $\vec{r}$ is the distance vector between the two dipoles, $r = |\vec{r}|$, and $\vartheta$ is the angle between the $z$-axis and $\vec{r}$.

 In the mean-field approximation, the two dipolar BECs are described by two coupled time-dependent GP equations~\cite{yi00,gora02}
\begin{eqnarray}
\label{eq_gp1}
i \hbar \frac{\partial \Psi_j(\vec{r},t)}{\partial t} =
\bigg[
H_j^0(\vec{r}) + \nonumber  \\
(N_j-1)  \int d^3 \vec{r}\,'|\Psi_j(\vec{r}\,',t)|^2  V_{\mathrm{dd}}(\vec{r}-\vec{r}\,') + \nonumber  \\
N_l  \int d^3 \vec{r}\,' |\Psi_l(\vec{r}\,',t)|^2 V_{\mathrm{dd}}(\vec{r}-\vec{r}\,')
 \bigg]
\Psi_j(\vec{r},t),
\end{eqnarray}
where $(j,l) = (1,2)$ and $(2,1)$, and $H_j^0$ denotes the single particle Hamiltonian,
\begin{eqnarray}
\label{eq_fham1}
H_j^0(\vec{r}) = - \frac{\hbar^2}{2m}\nabla^2 + V_{tj}(\rho,z).
\end{eqnarray}
The coupling between the wave functions $\Psi_1(\vec{r},t)$ and $\Psi_2(\vec{r},t)$ arises due to the dipole-dipole interaction between the two clouds, which is accounted for by the third term in the square bracket on the right hand side of Eq.~(\ref{eq_gp1}). The coupled mean field equations depend on four parameters, the aspect ratio $\lambda$, the separation $b$, and the dimensionless dipole strengths $D_1$ and $D_2$, where
\begin{equation}
\label{eq_dmf}
D_j= \frac{d^2(N_j-1)}{E_{\rho} a_{\rho}^3}.
\end{equation}
Here, $E_{\rho}$ and $a_{\rho}$ denote the oscillator energy and length along the $\rho$-direction, $E_{\rho} = \hbar \omega_{\rho}$ and $a_{\rho} = \sqrt{\hbar/(m \omega_{\rho})}$. The wave functions $\Psi_j$ are normalized according to $\int d^3 \vec{r} |\Psi_j(\vec{r},t)|^2 = 1$. Since the confining potential and the dipole-dipole interaction potential are cylindrically symmetric, we can write the wave functions as $\Psi_j(\vec{r}, t) = \widetilde\Psi_j(\rho,z, t) \exp(i k \phi)$, where $k = 0,\pm 1,\pm 2, \cdot\cdot\cdot$. 

The ground state solution of Eq.~(\ref{eq_gp1}) can be obtained by solving the  coupled time-independent GP equations self-consistently~\cite{gora02}. To this end, we set $k=0$ and write $\Psi_j(\vec{r},t)= \psi_j^0(\rho,z) \exp(-i \mu_j t/\hbar)$, where the $\mu_j$ denote the chemical potentials corresponding to the ground state solutions $\psi_j^0(\rho,z)$. We solve the  coupled time-independent GP equations self-consistently by evolving an initial state in imaginary time until convergence is reached~\cite{bao03}. Our numerical implementation exploits the cylindrical symmetry of the system and uses a two-dimensional grid in the $\rho$- and $z$-directions~\cite{rone06,asad09}.

In addition to the stationary ground state wave functions $\psi_j^0$ and the chemical potentials $\mu_j$, we determine the total energy per particle $E_{tot}/N$,
\begin{eqnarray}
\label{eq_etot}
E_{tot} / N = \sum_{j=1,2}\frac{N_j}{N}\int d^3 \vec{r} \big( \psi_j^{0}(\vec{r})\big)^\ast \bigg[ - \frac{\hbar^2}{2m}\nabla^2 + V_{tj}(\rho,z) + \nonumber\\
\frac{N_j-1}{2}\int d^3 \vec{r}\,' |\psi_j^0(\vec{r}\,')|^2  V_{\mathrm{dd}}(\vec{r} -\vec{r}\,')\bigg] \psi_j^0(\vec{r}) +\nonumber\\
\frac{N_1 N_2}{N}\int d^3 \vec{r}\int d^3 \vec{r}\,' |\psi_1^0(\vec{r}\,')|^2  V_{\mathrm{dd}}(\vec{r} -\vec{r}\,') |\psi_2^0(\vec{r})|^2,
\end{eqnarray}
where $N=N_1+N_2$. Here, we used $\psi_j^0(\rho,z) = \psi_j^0(\vec{r})$ for notational convenience. The first, second, and third terms in the square bracket on the right side of Eq.~(\ref{eq_etot}) give rise to the kinetic energy per particle $E_{kin} / N$, the trap energy per particle $E_{trap} / N$, and the on-site dipole-dipole interaction energy per particle $E_{dd,on} / N$, respectively. The off-site dipole-dipole interaction energy per particle $E_{dd,off} / N$ is given by the last term on the right hand side of Eq.~(\ref{eq_etot}).

We note that the system considered here can be viewed as a variant of the first mean-field study of two-component dipolar BECs by G\'oral \emph{et al.}~\cite{gora02} who considered the limiting case of vanishing separation and spherical confinement. However, as opposed to two dipolar BECs aligned along the same direction as considered here, they considered two oppositely polarized BECs. We have checked for selected cases that our solutions for the coupled stationary GP equations agree with those reported by G\'oral \emph{et al.}

\subsection{Coupled Bogoliubov de Gennes equations}
\label{sec_bdg}

To analyze the dynamical behavior of the system, we write~\cite{dalf99,dalf97}
\begin{eqnarray}
\label{eq_bdg1}
\Psi_j(\vec{r},t) = \left[ \psi_j^0(\vec{r})+
\delta \psi_j(\vec{r},t) \right] \exp(-i \mu_j t / \hbar),
\end{eqnarray}
where $j =1$ and $2$, and the $\delta \psi_j(\vec{r},t)$ denote the perturbation of the  dipolar BEC located in trap $j$. Following the literature~\cite{dalf97}, we write the perturbations in terms of the Bogoliubov particle and hole excitations $u_j$ and $v_j$, 
\begin{eqnarray}
\label{eq_per1}
\delta \psi_j(\vec{r},t) = 
u_j(\vec{r}) \exp(-i \omega t) 
+ v_j^*(\vec{r}) \exp( i \omega t).
\end{eqnarray}
Plugging Eqs.~(\ref{eq_bdg1}) and ~(\ref{eq_per1}) into Eq.~(\ref{eq_gp1}) and keeping terms up to the first order in $u_j$ and $v_j$, we find, after equating the coefficients of $\exp(i\omega t)$ and $\exp(-i\omega t)$, a set of two coupled BdG equations,
\begin{eqnarray}
\label{eq_bdf1}
\hbar^2 \omega^2 f_j(\vec{r}) = 
{\cal{A}}_j(\vec{r}){\cal{A}}_j(\vec{r}) f_j(\vec{r}) + 
\nonumber \\
{\cal{A}}_j(\vec{r}) \bigg[ 2 (N_j-1)\int d^3\vec{r}\,' \psi_j^0(\vec{r}\,')f_j(\vec{r}\,') V_{\mathrm{dd}}(\vec{r}-\vec{r}\,')  + 
\nonumber \\
2 N_l \int d^3\vec{r}\,'\psi_l^0(\vec{r}\,')f_l(\vec{r}\,')  V_{\mathrm{dd}}(\vec{r}-\vec{r}\,') 
\bigg] \psi_j^0(\vec{r}),
\end{eqnarray}
where, as before, $(j,l) = (1,2)$ and ($2,1$). In deriving Eq.~(\ref{eq_bdf1}), we assumed that the $\psi_j^0(\vec{r})$ are real. The functions $f_j(\vec{r})$, $ f_j(\vec{r}) = u_j(\vec{r}) + v_j(\vec{r})$, represent the density perturbation for the dipolar BEC located in the $j^{th}$ trap~\cite{rone06,pita03}. This becomes clear if we calculate the density using Eq.~(\ref{eq_bdg1}). Assuming that the $u_j$ and $v_j$ are real and keeping only the lowest order correction, we obtain $\big|\Psi_j(\vec{r},t)\big|^2 \approx \big|\psi_j^0(\vec{r})\big|^2 + 2 \cos(\omega t)\psi_j^0(\vec{r}) f_j(\vec{r})$. Because of the cylindrical symmetry of the problem, the density perturbations or eigen modes $f_j(\vec{r})$ can be written as $\bar{f}_j(\rho,z)\exp(ik\phi)$,  $k = 0,\pm 1,\pm 2, \cdot\cdot\cdot$~\cite{peth02}. The operators ${\cal{A}}_j(\vec{r})$ in Eq.~(\ref{eq_bdf1}) operate on everything to their right and are  given by
\begin{eqnarray}
\label{eq_capa1}
{\cal{A}}_j(\vec{r}) = H_j^0(\vec{r}) + (N_j -1) \int d^3\vec{r}\,' |\psi_j^0(\vec{r}\,')|^2  V_{\mathrm{dd}}(\vec{r}-\vec{r}\,') + \nonumber\\
 N_l \int d^3\vec{r}\,' |\psi_l^0(\vec{r}\,')|^2 V_{\mathrm{dd}}(\vec{r}-\vec{r}\,')-\mu_j.
\end{eqnarray}

We solve Eq.~(\ref{eq_bdf1}) for the eigen frequencies $\omega$ and the density perturbations $f_1$ and $f_2$ for various $k$, $k =0-4$, using the Arnoldi method~\cite{arno51}. Our implementation follows that discussed in Ref.~\cite{rone06} for a single component dipolar BEC. In particular, we construct a vector from the density perturbations $f_1$ and $f_2$, and then proceed as in the single component case. The excitation frequencies $\omega$ allow for the determination of the dynamical stability of the system. A positive $\omega^2$, or real $\omega$, signals that the system is dynamically stable with respect to the associated density oscillation. A negative $\omega^2$, or imaginary $\omega$, in contrast, signals that the system is dynamically unstable with respect to the associated density oscillation. As detailed further in Sec.~\ref{sec_results}, the dynamical instability of the two well dipolar system can, depending on the system parameters, be triggered by either a $k=0$ mode or a finite $k$ mode.

\section{Results}
\label{sec_results}

\begin{figure}
\vspace*{.2cm}
\includegraphics[angle=0,width=70mm]{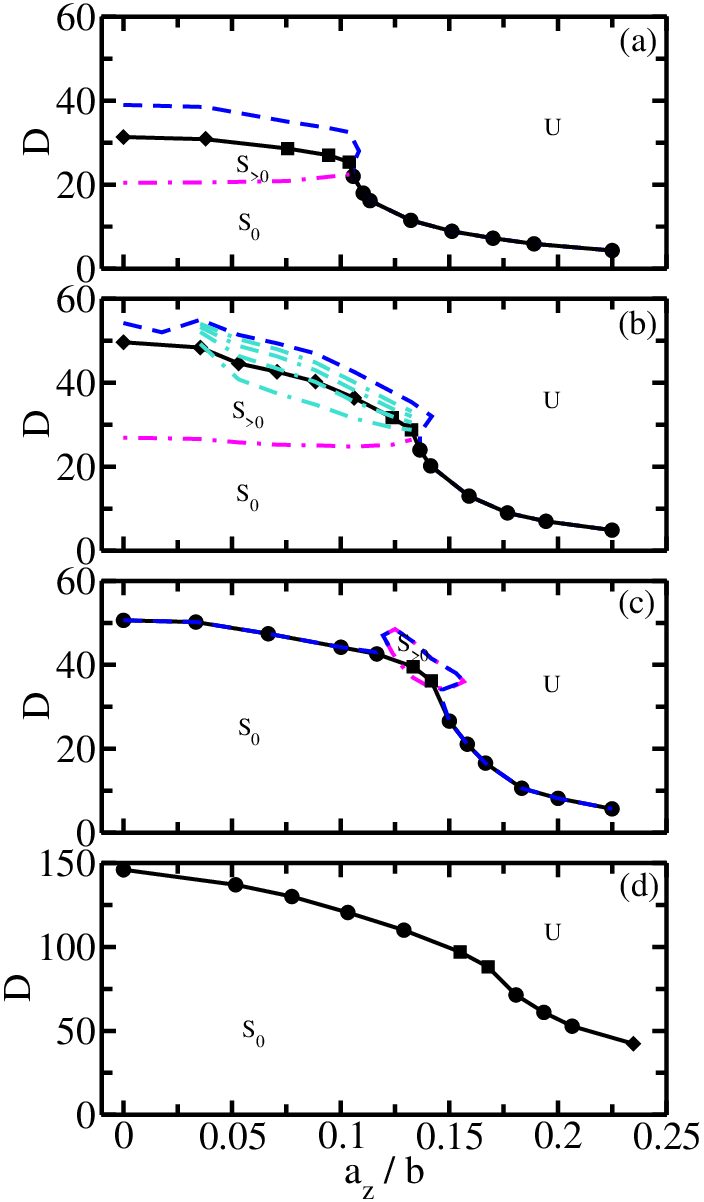}
\caption{
(Color online)
$D$-versus-$1/b$ phase diagram for (a) $\lambda = 7$, (b) $\lambda = 8$, (c) $\lambda = 9$, and (d) $\lambda = 15$. The solid lines separate the dynamically stable region from the dynamically unstable region (labeled as ``U''). Circles, squares and diamonds indicate that the system collapses through the softening of a $k = 0$, $k = 2$ and $k = 3$ excitation mode, respectively. In panels (a), (b), and (c), the dashed lines separate the mechanically stable region from the mechanically unstable region while  the dash-dotted lines mark the boundary between regions where density maxima are located at $\rho = 0$ (labeled as ``$\mathrm{S}_0$'') and where density maxima are located away from $\rho = 0$ (labeled as ``$\mathrm{S}_{>0}$''). For $\lambda = 8$, panel (b), dash-dash-dotted lines indicate, from bottom to top, $10\%$,  $20\%$,  $30\%$, and  $40\%$ difference between $n_j(0)$ and $n_j(\rho_{max})$ (see text). The oscillator length $a_z$ is defined through $a_z=\sqrt{\hbar/(m\omega_z)}$.
}\label{fig_phase}
\end{figure}

As discussed in Sec.~\ref{sec_theory}, the coupled GP equations depend on four parameters. To reduce the parameter space, we set $N_1=N_2=N/2$, and investigate the system properties as a function of $\lambda$, $b$, and $D$ ($D=D_1=D_2$). Figures~\ref{fig_phase}(a)-(d)  show the $D$-versus-$1/b$ phase diagram for $\lambda = 7, 8, 9$ and $15$, respectively. The solid lines separate the dynamically stable region from the dynamically unstable region. The symbols indicate the mode through which the system becomes unstable; circles, squares and diamonds correspond to the $k = 0, 2$ and $3$ mode, respectively. For a fixed separation $b$, the dynamically stable region increases with increasing $\lambda$.
This behavior is well known for a single dipolar BEC in a pancake-shaped trap~\cite{yi00,sant00}. As the aspect ratio $\lambda$ increases, the dipole-dipole interaction becomes effectively more repulsive. Figure~\ref{fig_phase} shows that the dynamically stable region decreases with decreasing separation, i.e., increasing $1/b$, for fixed $\lambda$. This decrease of stability is attributed to the presence of the second cloud. Since the dipoles are aligned along the $z$-axis, the dipole-dipole interaction between the two neighboring clouds, $E_{dd,off}$, becomes more attractive as the separation between the clouds is decreased.

The critical dipole strength $D_{cr}$, defined as the $D$ value for which the system becomes dynamically unstable, changes in an interesting manner with increasing $a_z/b$. For small $a_z/b$, $D_{cr}$ varies slowly. Around $a_z/b \gtrsim 0.11, 0.13, 0.14,$ and $0.17$, $D_{cr}$
\begin{figure}
\vspace*{.2cm}
\includegraphics[angle=0,width=70mm]{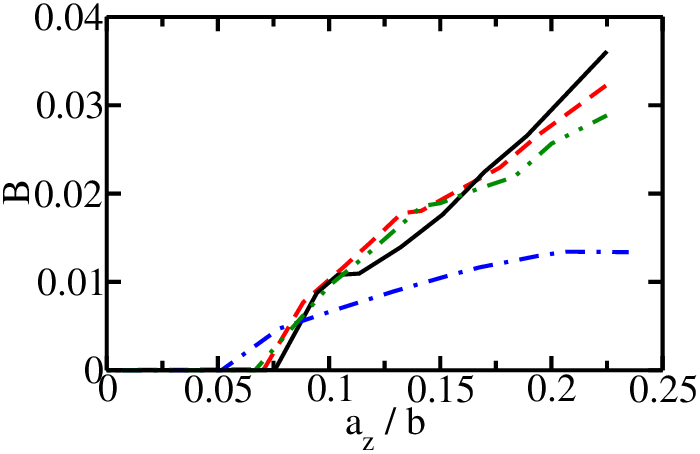}
\caption{
(Color online)
Asymmetry parameter $B$ as a function of $1/b$ for $\lambda = 7$ (solid line), $\lambda = 8$ (dashed line), $\lambda = 9$ (dot-dot-dashed line), and $\lambda = 15$ (dash-dotted line). For a given $\lambda$ and $b$, the $D$ value is chosen to be approximately equal to $0.9 D_{cr}$.
}\label{fig_asym}
\end{figure}
 varies comparatively fast for  $\lambda = 7, 8, 9$, and $15$, respectively. Finally, for larger $a_z/b$, $D_{cr}$ varies again comparatively slowly. In order to understand this dependence of $D_{cr}$ on $a_z/b$, we analyze the ground state density of the system. 

Dashed lines in Figs.~\ref{fig_phase}(a)-(c) show the mechanical instability line. The mechanical and dynamical instability lines nearly coincide when the ground state densities $|\psi_j^0|^2$ are approximately Gaussian shaped (labeled as  ``$\mathrm{S}_0$'') but deviate when the densities $|\psi_j^0|^2$ have a so-called red blood cell type shape (labeled as ``$\mathrm{S}_{>0}$''), i.e., when the density maxima are located at $\rho> 0$. The dash-dotted lines separate the two types of densities, which are determined by analyzing the integrated densities $n_j(\rho)$, where  $n_j(\rho)=2 \pi \int |\psi_{j}^0(\vec{r})|^2 dz$ and  $j=1$ and $2$, along the $\rho$ direction. If $n_j(0)/n_j(\rho_{\mathrm{max}})\leq 0.98$, where $\rho_{\mathrm{max}}$ is the $\rho$ value at which the integrated density $n_j(\rho)$ has its maximum, then we call the density red blood cell shape; otherwise we call it Gaussian. Figures~\ref{fig_phase}(a) and (b) show that the ground state density near the instability line has red blood cell type structure for fairly large separation. When the inverse separation $a_z/b$ has increased to about $0.1-0.13$, the red blood cell type structure disappears. $D_{cr}$ changes more rapidly for intermediate $a_z/b$ values when the ground state density is Gaussian. Figures~\ref{fig_phase}(a) and (b) suggest that the deformation of the ground state density away from the simple Gaussian like profile leads to a significant stabilization of the system. This interpretation is supported by the fact that the deformation, or the red blood cell type structure, becomes comparatively more pronounced as $a_z/b$ increases from $0$ to about $0.13$, as indicated by the dash-dash-dotted lines in Fig.~\ref{fig_phase}(b).
\begin{figure}
\vspace*{.2cm}
\includegraphics[angle=0,width=70mm]{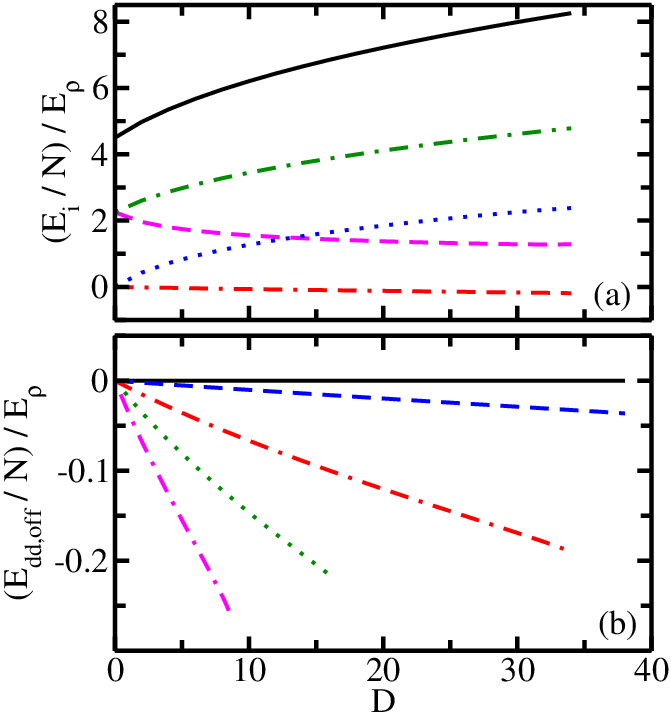}
\caption{
(Color online)
(a) Energy contributions $E_i/N$ as a function of $D$ for $\lambda = 7$ and $a_z/b = 1/(5\sqrt{\lambda})\approx 0.076$ ($z_{10} \approx -6.61 a_z$ and $z_{20}\approx 6.61 a_z$).
The solid, dashed, dash-dotted, dotted, and dash-dash-dotted lines show $E_{tot}/N$, $E_{kin}/N$, $E_{trap}/N$, $E_{dd,on}/N$, and $E_{dd,off}/N$, respectively. (b) The energy contribution  $E_{dd,off}/N$ is shown as a function of $D$ for $\lambda = 7$ and $a_z/b = 0$ (solid line), $a_z/b = 1/(10\sqrt{\lambda})\approx0.038$ (dashed line), $a_z/b= 1/(5\sqrt{\lambda})\approx0.076$ (dash-dash-dotted line), $a_z/b = 1/(3.33\sqrt{\lambda})\approx0.113$ (dotted line), and $a_z/b = 2/(5\sqrt{\lambda})\approx 0.151$ (dash-dotted line), respectively. 
}\label{fig_energy}
\end{figure}
 For larger aspect ratios [see Fig.~\ref{fig_phase}(c)], the ground state density in the dynamically stable region has red blood cell type structure only in a tiny region in the vicinity of the instability line around $a_z/b\approx 0.13$. Even so, for slightly larger $a_z/b$ values, $D_{cr}$ varies more rapidly. For $\lambda = 15$ [see Fig.~\ref{fig_phase}(d)], the red blood cell type structure exists in an even smaller region around $(D,a_z/b)\approx (90,0.17)$, where the system collapses through a $k=2$ mode. Although much less pronounced, $D_{cr}$ varies more rapidly for slightly larger $a_z/b$ values. Interestingly, the density deviates again from the simple Gaussian type shape in a tiny region around $(D,a_z/b)\approx (35,0.235)$

To further characterize the ground state density, we define the quantity $B_j$, which measures the asymmetry of the density of the cloud located in the $j^{th}$ trap, i.e., the density asymmetry about the trap center,
\begin{eqnarray}
\label{eq_asym1}
B_j = 2 \pi \bigg[\int_0^\infty \int_{-\infty}^{z_{j0}} \rho d\rho dz |\psi_j^0(\vec{r})|^2 -\nonumber\\
\int_0^\infty\int_{z_{j0}}^{\infty} \rho d\rho dz  |\psi_j^0(\vec{r})|^2
\bigg],
\end{eqnarray}
where $j=1$ and $2$. For equal number of dipoles in both traps, as considered in this paper, one has $B_1 = -B_2$ and we define $B=|B_j|$. When the separation $b$ is large, the ground state densities $|\psi_j^0|^2$ are symmetric and $B=0$. However, as $b$ decreases, the interaction between the dipoles located in the two traps leads to an increased density between the trap centers and thus to finite $B$ values. To quantify how $B$ changes with decreasing $b$, we move along a ``trajectory'' in the $D$-versus-$1/b$ phase diagram for fixed $\lambda$ (see Fig.~\ref{fig_phase}) that lies $10\%$ below the solid line, i.e., we choose $D\approx 0.9 D_{cr}$ for fixed $\lambda$ and $b$. Figure~\ref{fig_asym} shows that the asymmetry parameter $B$ is, to within our numerical accuracy, identically zero for $a_z/b \lesssim 0.05-0.075$ for the trajectories investigated. Figure~\ref{fig_asym} shows that it takes a certain critical attraction before the system breaks the symmetry of the ground state density. For large separations, the energy is minimized for densities symmetric about $z_{j0}$. For smaller $b$, however, the off-site interaction is attractive enough to deform the ground state densities $|\psi_j^0|^2$.

To gain further insight, Fig.~\ref{fig_energy}(a) shows the energy per particle, $E_{tot}/N$, as well as the individual energy contributions as a function of $D$ for $\lambda = 7$ and $a_z/b \approx 0.076$. $E_{tot} / N$ (solid line), $E_{trap} / N$ (dash-dotted line) and $E_{dd,on} / N$ (dotted line) increase  with increasing $D$ while $E_{kin} / N$ (dashed line) and  $E_{dd,off} / N$ (dash-dash-dotted line) decrease with increasing $D$. Since the dipoles are aligned along the $z$-direction and the system is pancake-shaped, $E_{dd,on}$ is effectively repulsive. To minimize $E_{dd,on}$, the dipoles try to spread out, which reduces $E_{kin}$ but increases $E_{trap}$ as $D$ increases. $E_{dd,off}$ decreases (i.e., becomes more negative) with increasing $D$ since the attraction between the dipoles located in the two traps increases. This qualitative behavior remains the same as the separation $b$ between the traps decreases. For fixed $D$ and $\lambda$, the energy contribution that changes the most is $E_{dd,off}$. As shown in Fig.~\ref{fig_energy}(b), $E_{dd,off}/N$ decreases appreciably as the separation $b$ decreases, which can be attributed to the attraction between dipoles located in the two different clouds. The fact that $E_{dd,off}/N$ is the energy contribution that changes the most as $b$ decreases emphasizes that the decreased stability is driven by the dipole-dipole interactions between the two clouds.

We now investigate how the collapse mechanism changes with $b$.
\begin{figure}
\vspace*{.2cm}
\includegraphics[angle=0,width=70mm]{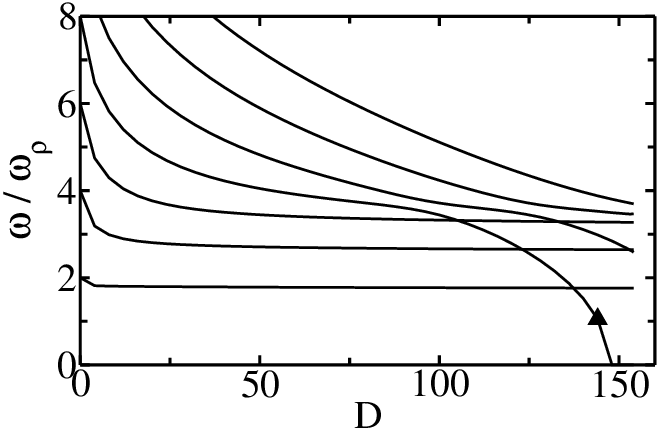}
\caption{
Solid lines show the seven lowest non-zero BdG 
excitation frequencies $\omega$ as a function of $D$ for 
$\lambda = 15$, $a_z/b = 0$, and $k = 0$. 
The triangle marks the BdG eigen frequency for which 
Figs.~\ref{fig_eigen_bin_0} and \ref{fig_psisq_bin_0} show the corresponding 
eigen mode $\bar{f}_1(\rho,z)$ and ground state density, respectively.
}\label{fig_exci_m_0_lam_15_bin_0}
\end{figure}
For infinite separation, each BdG excitation frequency is doubly degenerate and the system collapses through a $k=0$ mode when the ground state density has Gaussian shape and through a $k>0$ mode when the ground state density has red blood cell shape~\cite{rone07}. Figure~\ref{fig_exci_m_0_lam_15_bin_0} shows the seven lowest BdG eigen frequencies as a function of $D$ for $k=0$, $\lambda=15$, and $a_z/b=0$. In this case, the system collapse is triggered by a radial roton mode.
\begin{figure}
\vspace*{.2cm}
\includegraphics[angle=0,width=70mm]{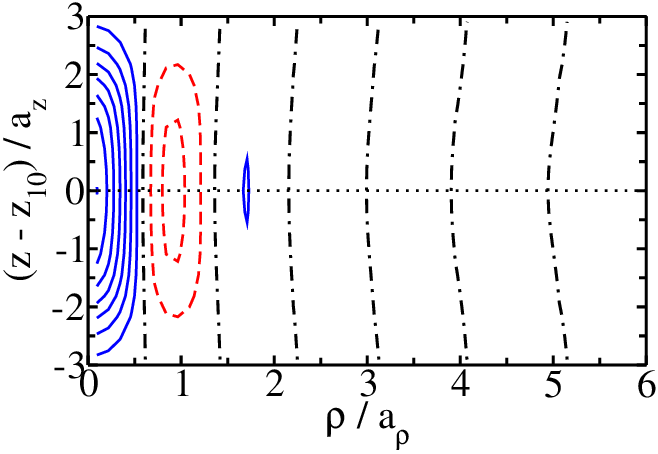}
\caption{
(Color online)
Contour plot of the BdG eigen mode $\bar{f}_1(\rho, z)$ for $\lambda = 15$, $a_z/b = 0$, $D \approx 144$, and $k=0$. The contours are equally spaced. Solid and dashed lines indicate positive and negative $\bar{f}_1(\rho,z)$ values, respectively. The dash-dotted lines indicate the nodal lines and the dotted line marks the trap center.
}\label{fig_eigen_bin_0}
\end{figure}
Figure~\ref{fig_eigen_bin_0} shows the eigen mode $\bar{f}_1$ corresponding to the lowest BdG eigen frequency for $D\approx 144$ (see triangle in Fig.~\ref{fig_exci_m_0_lam_15_bin_0}). The density oscillation $\bar{f}_1$ has six nodal lines that are separated by approximately $0.8 a_{\rho}$. This nodal line spacing agrees quite well with $\lambda_{\rho}/2$, where $\lambda_{\rho}$ is the wavelength expected for a radial roton mode, $\lambda_{\rho}\approx2\pi a_z\approx1.62 a_{\rho}$~\cite{rone07,sant03}. 

\begin{figure}
\vspace*{.2cm}
\includegraphics[angle=0,width=70mm]{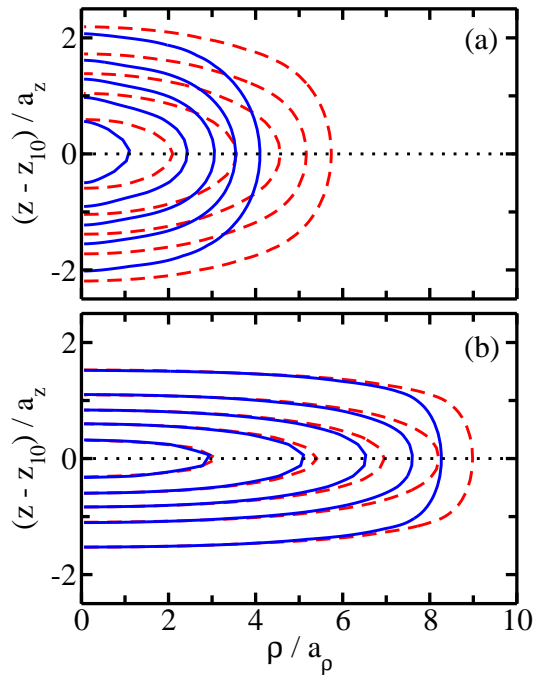}
\caption{
(Color online)
Contour plot of the ground state density  for $\lambda = 15$
calculated (a) 
accounting for the full dynamics in the 
$z$-direction and (b) assuming frozen dynamics 
in the $z$-direction.
The solid lines correspond to
$a_z/b \approx 0.21$ and (a) $D \approx 52$ and (b) $D \approx 512$.
The dashed lines correspond to
$a_z/b=0$ and (a) $D \approx 144$ and (b) $D \approx 608$.
The contours correspond to $90\%$ (centermost contour), 
$70\%$, $50\%$, $30\%$ and
$10\%$ (outermost contour) of the peak density.
The dotted lines mark the trap center.
}\label{fig_psisq_bin_0}
\end{figure}

As $b$ decreases, the collapse mechanism of the system 
\begin{figure}
\vspace*{.2cm}
\includegraphics[angle=0,width=70mm]{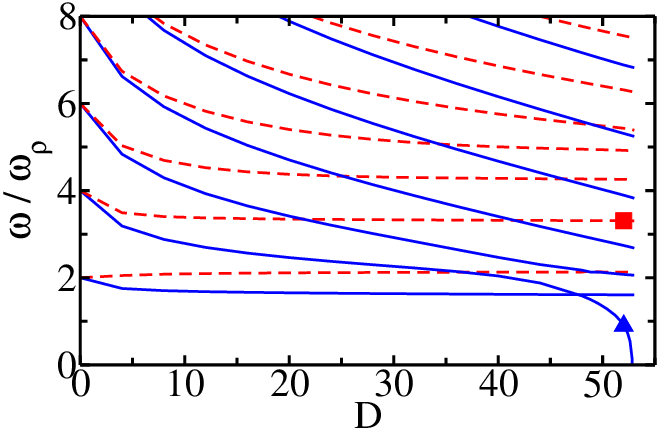}
\caption{
(Color online)
Solid and dashed lines show the in-phase and 
out-of-phase BdG excitation frequencies $\omega$ 
as a function of $D$ for $\lambda = 15$, 
$a_z/b=4/(5\sqrt{\lambda})\approx 0.21$ 
($z_{10} \approx -2.42 a_z$ and $z_{20} \approx 2.42 a_z$), 
and $k = 0$. 
The triangle and square mark the BdG eigen 
frequencies for which Fig.~\ref{fig_eigen_bin_p8_out} shows 
the corresponding eigen modes $\bar{f}_1(\rho,z)$ and $\bar{f}_2(\rho,z)$;
the corresponding ground state density is shown in Fig.~\ref{fig_psisq_bin_0}.
}\label{fig_exci_m_0_lam_15_bin_p8}
\end{figure}
changes due to the attractive dipole-dipole interaction between the two clouds.
\begin{figure}
\vspace*{.2cm}
\includegraphics[angle=0,width=60mm]{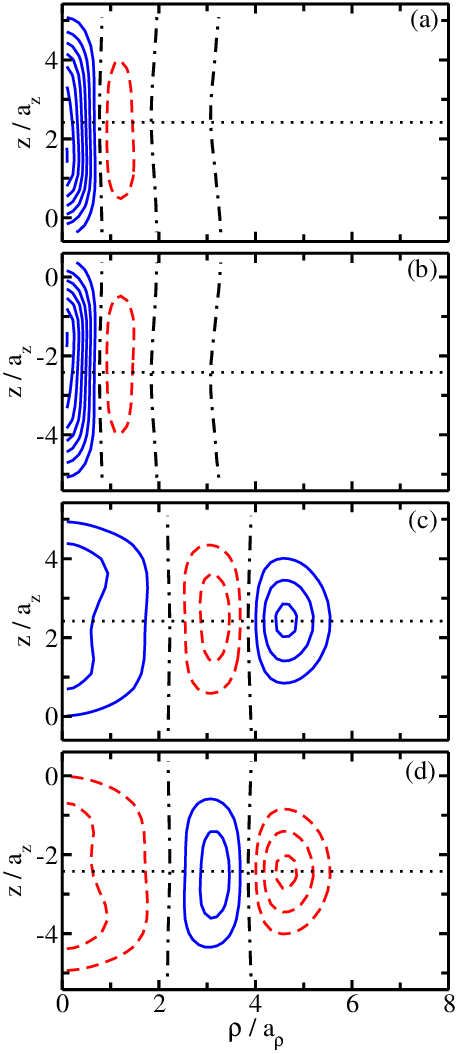}
\caption{
(Color online)
Contour plot of (a) and (b) the in-phase BdG eigen modes $\bar{f}_1(\rho,z)$ and $\bar{f}_2(\rho,z)$, respectively, and (c) and (d) the out-of-phase BdG modes $\bar{f}_1(\rho,z)$ and $\bar{f}_2(\rho,z)$, respectively, for $\lambda = 15$,  $a_z/b=4/(5\sqrt{\lambda})\approx 0.21$ ($z_{10} \approx -2.42 a_z$ and $z_{20} \approx 2.42 a_z$), $D \approx 52.02$, and $k =0$. The contours are equally spaced. Solid and dashed lines indicate positive and negative $\bar{f}_1(\rho,z)$ and $\bar{f}_2(\rho,z)$ values, respectively. The dash-dotted lines indicate the nodal lines and the dotted lines mark the trap center. The eigen frequency corresponding to panels (a) and (b) is marked by a triangle in Fig.~\ref{fig_exci_m_0_lam_15_bin_p8} while that corresponding to panels (c) and (d) is marked by a square.
}\label{fig_eigen_bin_p8_out}
\end{figure}
Figure~\ref{fig_exci_m_0_lam_15_bin_p8} shows the fourteen lowest BdG eigen frequencies as a function of $D$ for $k=0$, $\lambda=15$, and $a_z/b\approx 0.21$. Solid and dashed lines indicate that the BdG eigen frequencies correspond to in-phase and out-of-phase density oscillations of the BECs located in the two traps. For $D=0$, the eigen spectrum consists of degenerate pairs. As $D$ increases, the in-phase and out-of-phase frequency pairs decouple. This behavior is analogous to that of a symmetric one-dimensional double-well potential. In the weak coupling regime (high barrier), the tunneling splitting is small and the eigen spectrum consists of nearly degenerate pairs. As the coupling increases, the eigen frequency pairs decouple. Figure~\ref{fig_exci_m_0_lam_15_bin_p8} shows that the lowest in-phase eigen frequency approaches zero for $D\approx 52.9$. Figures~\ref{fig_eigen_bin_p8_out}(a) and (b) show the corresponding eigen modes for a slightly smaller $D$ value 
(see triangle in Fig.~\ref{fig_exci_m_0_lam_15_bin_p8}; 
the corresponding ground state
density is shown in Fig.~\ref{fig_psisq_bin_0}). 
As a result of the attractive off-site dipole-dipole interaction, the in-phase eigen modes $\bar{f}_1$ [Fig.~\ref{fig_eigen_bin_p8_out}(a)] and $\bar{f}_2$ [Fig.~\ref{fig_eigen_bin_p8_out}(b)] just prior to collapse are slightly asymmetric around the trap centers $z_{10}$ and $z_{20}$, respectively. The in-phase eigen modes $\bar{f}_1$ and $\bar{f}_2$ have three nodal lines, whose separation increases slightly with increasing $\rho$. This suggests that the coupled dipolar BEC system does not, like the uncoupled system (see Figs.~\ref{fig_exci_m_0_lam_15_bin_0} and~\ref{fig_eigen_bin_0}), collapse through a ``pure'' radial roton mode. The radial roton mode can be interpreted as being the result of the formation of a pattern along the $\rho$-direction whose size is governed by $a_z$. In the presence of the second dipolar BEC, the separation $b$ between the two clouds sets another length scale. For the parameters in Figs.~\ref{fig_exci_m_0_lam_15_bin_p8} and~\ref{fig_eigen_bin_p8_out}, we have $b = 1.25 a_{\rho}\approx 4.84 a_z$. Thus, the dynamics along the $\rho$-direction is governed by an interplay of the length scales $a_z$ and $b$,  resulting in modes $\bar{f}_1$ and $\bar{f}_2$ that have neither the characteristic features of a ``pure'' radial roton mode nor those of a ``pure'' breathing mode of the entire two-cloud system.

For comparison, Figs.~\ref{fig_eigen_bin_p8_out}(c) and (d) show the eigen modes $\bar{f}_1$ and $\bar{f}_2$ corresponding to the second lowest out-of-phase frequency for $D\approx 52.02$ (see square in Fig.~\ref{fig_exci_m_0_lam_15_bin_p8}). For vanishing $D$, the corresponding eigen frequency is degenerate with the in-phase eigen frequency that, for finite $D$, triggers the collapse. The nodal pattern of the out-of-phase eigen modes [Figs.~\ref{fig_eigen_bin_p8_out}(c) and (d)] is distinctly different from that of the in-phase eigen modes [Figs.~\ref{fig_eigen_bin_p8_out}(a) and (b)], underlining the fact that the eigen frequency  pairs decouple as the coupling between the clouds increases.

\begin{figure}
\vspace*{.2cm}
\includegraphics[angle=0,width=70mm]{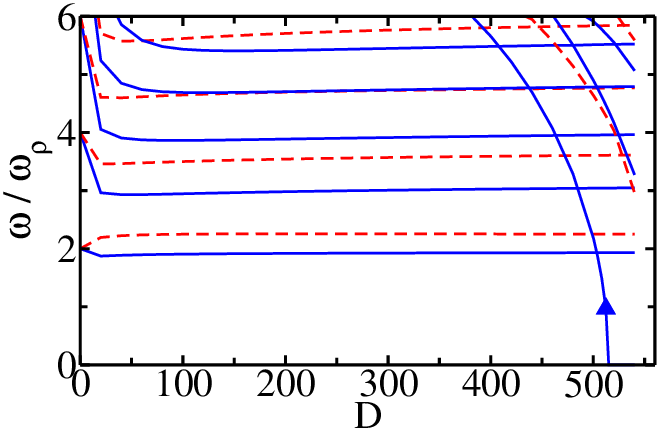}
\caption{
(Color online)
Solid and dashed lines show the in-phase and out-of-phase
BdG 
excitation frequencies $\omega$ obtained assuming that the $z$-dynamics is frozen as a function 
of $D$ for $\lambda = 15$, 
$a_z/b \approx 0.21$, 
and $k=0$. 
The triangle marks the BdG eigen frequency for which Figs.~\ref{fig_psisq_bin_0} and ~\ref{fig_eigen_fbin_0} show the corresponding density and eigen mode $\bar{f}_1(\rho,z)$, respectively.
}\label{fig_exci_m_0_lam_15_fbin_0}
\end{figure}

\begin{figure}
\vspace*{.2cm}
\includegraphics[angle=0,width=70mm]{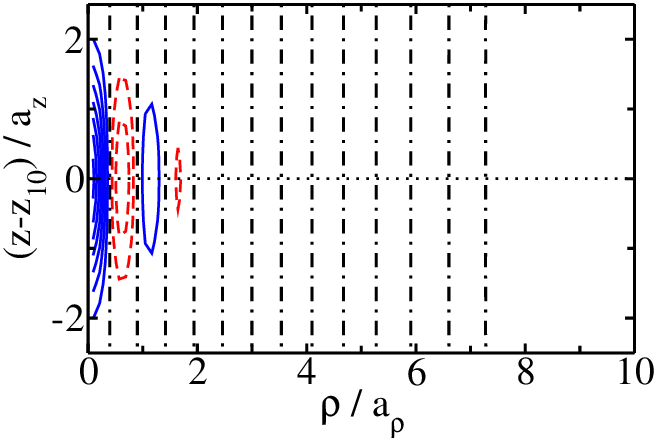}
\caption{
(Color online)
Contour plot of the BdG eigen mode $\bar{f}_1(\rho, z)$ 
for $\lambda = 15$, 
$a_z/b \approx 0.21$, 
$D \approx 512$, 
and $k=0$ obtained assuming that the $z$-dynamics is frozen. 
The contours are equally spaced. Solid and 
dashed lines indicate positive and negative $\bar{f}_1(\rho, z)$ values, respectively. The dash-dotted lines indicate the nodal lines and the dotted line marks the trap center. The eigen frequency corresponding to this eigen mode is marked by a triangle in Fig.~\ref{fig_exci_m_0_lam_15_fbin_0}.
}\label{fig_eigen_fbin_0}
\end{figure}

Finally, we investigate the system behavior assuming that the dynamics in the $z$-direction is frozen, i.e., we write~\cite{wang08,kobe09,klaw09}
\begin{equation}
\label{eq_anst}
\Psi_j(\vec{r},t) = \widetilde{\psi}_j(\rho,\phi,t) \phi_j^0(z),
\end{equation}
where the $\phi_j^0(z)$ denote the one-dimensional harmonic 
oscillator ground state wave functions of 
the $j^{th}$ trap in the $z$-direction. 
For the parameter combinations investigated,
the ground state densities obtained using the frozen
$z$-dynamics approach show, just as the densities obtained using the 
full mean-field dynamics, 
Gaussian and red blood cell type structures
in the vicinity of the mechanical and dynamical instabilities. However, these
structures appear at different $(D,b)$ combinations for the two 
different approaches.
We find that the variational wave function 
given in Eq.~(\ref{eq_anst}) predicts the mechanical
instability to set in at much larger $D$ values
than predicted by the mean-field wave function
that accounts for the full dynamics.
Moreover, the frozen $z$-dynamics approach predicts a
fairly smooth decrease of $D_{cr}$ with increasing $1/b$ (for $a_z/b < 0.25$) and
does not reproduce the relatively steep drop of 
$D_{cr}$ around $a_z/b=0.1-0.175$ discussed in the context
of Fig.~\ref{fig_phase}.
Figures~\ref{fig_psisq_bin_0}-\ref{fig_eigen_fbin_0}
exemplarily illustrate these findings.

To start with, we analyze the energetics for
$\lambda = 15$, $a_z/b \approx 0.21$ and $D \approx 52$
(see triangle in Fig.~\ref{fig_exci_m_0_lam_15_bin_p8}). 
We find that the ground state energy obtained 
for the variational wave function, Eq.~(\ref{eq_anst}), 
is about $5\%$ higher than the exact mean-field energy. 
While the total energy agrees fairly well, 
the kinetic energy $E_{kin}$ differs by about
$40\%$, suggesting that the description based on the
frozen $z$-dynamics is not flexible enough 
to describe all features of the system qualitatively correctly. 
Indeed, the frozen $z$-dynamics approach
predicts the dynamical instability
to occur at $D_{cr} \approx 515$ (see
Fig.~\ref{fig_exci_m_0_lam_15_fbin_0}), i.e., $D_{cr}$
predicted by the frozen $z$-dynamics is about ten times larger
than $D_{cr}$ predicted by the full mean-field dynamics.

Solid lines in Figs.~\ref{fig_psisq_bin_0}(a) and (b) 
compare the contour plots
of the
ground state density
obtained accounting for the full dynamics 
and assuming frozen $z$-dynamics, respectively, for
$\lambda=15$, $a_z/b \approx 0.21$ and 
$D$
values that are slightly smaller
than the respective $D_{cr}$, 
i.e., $D \approx 52$ in Fig.~\ref{fig_psisq_bin_0}(a)
and 
$D \approx 512$ in Fig.~\ref{fig_psisq_bin_0}(b).
The ground state density obtained assuming frozen 
$z$-dynamics [Fig.~\ref{fig_psisq_bin_0}(b)] 
is significantly more extended in the $\rho$-direction and 
less extended in the $z$-direction than that obtained 
accounting for the full dynamics 
[Fig.~\ref{fig_psisq_bin_0}(a)]. 
For comparison, dashed lines in Figs.~\ref{fig_psisq_bin_0}(a) and (b)
show the ground state density for the same $\lambda$ value
(i.e., $\lambda=15$), but $1/b=0$. The $D$ values are, as for the densities
shown by solid lines, chosen to 
be slightly smaller than the respective
$D_{cr}$ values, i.e., $D \approx 144$ in Fig.~\ref{fig_psisq_bin_0}(a) 
and $D \approx 608$ in Fig.~\ref{fig_psisq_bin_0}(b).
Compared to the densities for finite separation, those for
infinite separation are more extended in the $\rho$ direction.

Figure~\ref{fig_exci_m_0_lam_15_fbin_0} shows 
the 
BdG eigen spectrum 
obtained assuming frozen dynamics in the $z$-direction 
as a function of $D$
for $k=0$, $\lambda = 15$, and $a_z/b \approx 0.21$. 
A comparison of Figs.~\ref{fig_exci_m_0_lam_15_fbin_0} and  
\ref{fig_exci_m_0_lam_15_bin_p8} shows 
that the 
spectrum obtained based on the frozen $z$-dynamics reproduces that
obtained based on the full dynamics qualitatively but not
quantitatively.
In particular, the frozen $z$-dynamics approach predicts 
$D_{cr} \approx 515$, compared to $D_{cr} \approx 52.9$ obtained using the full
dynamics approach.
Figure~\ref{fig_eigen_fbin_0} shows the eigen mode 
$\bar{f}_1$ for $\lambda = 15$, 
$a_z/b \approx 0.21$, $D \approx 512$ (which is just a bit
smaller than $D_{cr}$), and $k=0$ for the 
lowest eigen frequency. 
The eigenmode possesses 13 nodal lines, which are approximately
equally spaced (the spacing is about
$0.55 a_{\rho}$ for the first 8 or 9 nodal lines
and slightly larger for the last 5 or 4 nodal lines),
indicating that the collapse is triggered,
according to the frozen $z$-dynamics approach,
by a radial roton mode and not, as predicted by
the full mean-field dynamics, by a mode that is neither a
pure radial roton mode
nor a pure breathing mode
[see discussion around Fig.~\ref{fig_eigen_bin_p8_out}(a)].
Although it might be expected intuitively that an aspect ratio of $\lambda = 15$ is sufficiently large to treat the system as effectively one-dimensional, our analysis shows that this is not the case. 
Our study suggests that caution needs to be exercised when the dynamics of coupled pancake shaped traps is treated within a variational approach. Future studies need to extend the analysis to even higher $\lambda$ to make direct contact with Refs.~\cite{wang08,kobe09,klaw09}.  

\section{Summary}
\label{sec_conclusion}
We studied the behavior of two coupled dipolar BECs, each located in a cylindrically symmetric external confining potential, as the separation $b$ between the traps along the tight confining direction is varied. The dipoles are aligned along the $z$-direction and $s$-wave interactions are neglected. The number of dipoles in each trap is conserved separately, i.e., tunneling between the traps is neglected. The solutions of the coupled GP equations show that the system behavior is modified by the presence of the second dipolar BEC. As the separation is decreased from infinitely large values to a  value of about 
$5$ or $4 a_z$, 
initially the collapse behavior changes little and then significantly below a certain critical separation. For separations smaller than this critical separation, the presence of the second dipolar cloud destabilizes the system dramatically compared to the case where the traps are infinitely far apart. For certain parameter combinations, we find that the so called red blood cell type density becomes more pronounced or appears due to the presence of the second dipolar BEC. For infinitely large separation, each BdG frequency is doubly degenerate. As the separation is decreased, the solutions of the coupled BdG equations show that the eigen frequency pairs decouple into two eigen frequencies corresponding to in-phase and out-of-phase density oscillations of the BECs located in the two traps. When the separation between the traps is large, the system collapses through a radial roton mode if the ground state density is Gaussian shape and through an angular roton mode if the ground state density is red blood cell shape, similar to the case of a single dipolar BEC~\cite{rone07}. For relatively small separation, in contrast, the system collapses through a mode that is notably different from the radial roton mode that induces the collapse of a single dipolar BEC. For comparison, we also considered a simplified description in which the dynamics in the $z$-direction is assumed to be frozen. 
Compared to the full mean-field description, 
the simplified description, which is used 
frequently in the literature~\cite{wang08,kobe09,klaw09}, 
reproduces some features
qualitatively but not quantitatively.
We conclude that the frozen $z$-dynamics approach is inadequate to 
quantitatively
describe certain aspects of purely dipolar BECs, 
including the collapse, even if the aspect ratio is fairly large.

During the final stage of preparing this manuscript
for submission, we became aware of a related study
by Wilson and Bohn~\cite{wilson11} that
considers the dynamics of an array of dipolar
pancake-shaped BECs at various levels of approximation.

Support by the NSF through
grant PHY-0855332
is gratefully acknowledged.

\end{document}